\documentstyle[12pt]{article}
\newcommand{\ba}{\begin{array}}
\newcommand{\ea}{\end{array}}
\newcommand{\be}{\begin{equation}}
\newcommand{\ee}{\end{equation}}
\newcommand{\Nu}[2]{{ $^{#2}${#1} }}
\newcommand{\Dyu}{\Nu{Dy}{151} }
\newcommand{\Dyd}{\Nu{Dy}{152}}
\newcommand{\Gdz}{\Nu{Gd}{150}}
\newcommand{\Gdn}{\Nu{Gd}{149}}
\newcommand{\Tbu}{\Nu{Tb}{151}}
\newcommand{\Hgz}{\Nu{Hg}{190}}
\newcommand{\Hgd}{\Nu{Hg}{192}}

\newcommand{\Hgq}{\Nu{Hg}{194}}
\newcommand{\Pbq}{\Nu{Pb}{194}}

\newcommand{\ho}{\hbar\omega\,}

\newcommand{\Jd}{{\cal J}_2}

\title      {Microscopic study of superdeformation in the A=150 mass region}

\author     {
             P. Bonche \\
             {\em SPhT \thanks{Laboratoire de la DSM} -
              CE Saclay, 91191 Gif sur Yvette Cedex, France}
\and         H. Flocard \\
             {\em Division de Physique Th\'{e}orique\thanks
             {Unit\'e de recherches des Universit\'es Paris XI
             et Paris VI associ\'ee au CNRS.},
             Institut de Physique Nucl\'eaire,} \\
             {\em 91406 Orsay Cedex, France}
\and         P.-H. Heenen\thanks{Directeur de Recherches FNRS.} \\
             {\em Service de Physique Nucl\'{e}aire Th\'{e}orique,} \\
             {\em U.L.B - C.P. 229, B 1050 Brussels, Belgium}
            }

\date{\today}

\begin{document}

\maketitle

\newpage

\begin{abstract}
The cranked Hartree-Fock-Bogoliubov method presented in previous
studies of superdeformed bands in Hg and Pb isotopes is applied
to the study of superdeformed bands in \Gdz, \Dyd, \Dyu and \Tbu.
The same density-dependent zero-range pairing interaction
used in the A=190 mass region leads to
a similarly good agreement with the experimental data. In particular,
our results confirm the importance of a correct treatment
of pairing correlations in {\Gdz} to reproduce its experimental
dynamical moment of inertia.
The quasi-particle
spectra obtained for all the nuclei studied here are compatible with the
experimental results obtained for excited bands in this mass region.
The quality of our results opens the possibility of studying
microscopically very subtle phenomena like the properties
of identical bands.
\end{abstract}

\section{Introduction}

The $A$=150 mass region is the first
one where superdeformed (SD) bands have been
observed. The nucleus {\Dyd} at the center of this region benefits
optimally of the stability generated by the shell effects at
$N$=86 and $Z$=66. For large angular momenta, these
allow the formation of a SD well.
However, at low spins, quantal stabilization
is not strong enough to compensate
the increase of surface energy associated with the large quadrupole
deformation.
The physical situation is therefore
different from that encountered in the $A$=190 mass region
where one believes that SD bands disappear for values of spins
close to 10$\hbar$ or sometimes even smaller.
In {\Dyd} and neighbouring nuclei, the SD bands which have been detected
exist probably only at spins larger than 20$\hbar$.
In this angular momentum range, pairing correlations are
strongly weakened by the fast rotation and thus
expected to play a minor role. They become
necessary only for a detailed understanding of the data.

For this reason
the SD bands of this region can be classified within
models which neglect pairing correlations.
This was first attempted by
Bengtsson et al \cite{Ben88} using the Nilsson-Strutinsky approximation.
In particular, they showed that
the characteristic behavior of the second moment of inertia $\Jd$
of SD bands could be understood
in terms of the number of occupied
intruder orbitals (N=7 for neutrons and N=6 for protons).
There are indeed
close analogies between the frequency dependence
of the $\Jd$'s of SD bands of
different nuclei when they can be assigned
configurations with the same number of occupied intruder
orbitals.
Nazarewicz et al \cite{Naz89} have
used a phenomenological mean-field based on a Woods-Saxon potential
with pairing correlations treated within the Bogoliubov method
and shell effects calculated according to the Strutinsky procedure.
They have investigated the importance
of shape changes as well as that of pairing correlations.
and confirmed the adequacy of a classification of
SD bands according to the occupancy of intruder orbitals.
They have also shown that in some nuclei, like {\Gdz},
pairing correlations are necessary to explain
the frequency dependence of the dynamical moment
of inertia. Their study has been further extended within
the Woods-Saxon Strutinsky shell-correction method in the work of
Satu{\l}a et al \cite{Sat94} where
the self-consistency between
shape changes and dynamical pairing correlations is fully taken into account.
 This work also includes an approximate treatment of
particule number projection by means of the Lipkin-Nogami prescription.
To analyze the same self-consistency,
Shimizu et al \cite{Shi90} have used a model in which
pairing correlations are treated dynamically within
the RPA approximation.
All these studies indicate that among the nuclei
of this mass region, {\Gdz}
is one for which pairing should play an important role.

In a series of previous papers \cite{Gal94, Hee95, Ter95},
 we have presented a
cranked Hartree-Fock-Bogoliubov method in which pairing correlations
are treated dynamically thanks to the Lipkin-Nogami prescription.
A Skyrme force was used in the particle-hole channel and
a seniority interaction in the particle-particle channel\cite{Gal94, Hee95}.
The introduction of a density-dependent zero-range interaction
to describe pairing correlations
greatly improved the agreement with the experimental data\cite{Ter95}.
Transition energies were reproduced within 10 keV over
more than 10 transitions in the four isotopes {\Hgz}, {\Hgd}, {\Hgq}, {\Pbq}.
{\Gdz} provides an opportunity to test the properties of
this pairing interaction at much higer spins, in a regime
where pairing correlations are weakened.
For this reason, we have decided to study pairing effects
in this nucleus and in the magic SD nucleus {\Dyd}
where their effect should be
less important. We will also test the quality of
the neutron and proton single-particle spectra that
we obtain in two odd nuclei, {\Dyu} and {\Tbu}.

In the present study we have used a Skyrme effective
force with the SkM$^*$ parametrization
to describe the interaction between nucleons in the particle-hole channel.
In the particle-particle channel, several
treatments of pairing correlations have been tested.
First they have been neglected; in the following, the
corresponding Hartree-Fock (HF) results will
be refered to as (i).
Then, three types of Hartree-Fock-Bogoliubov (HFB) calculations have been
performed. In two of them, a seniority interaction
is used in the particle-particle channel. In one case
the strength is adjusted to reproduce the
two neutron separation energies
$( S_{2n} )$ at zero spin (ii).
The intensity of the other seniority force is smaller
and has been chosen to reproduce the
behaviour of the moment of inertia of {\Gdz} for $\ho\approx$500keV (iii).
The third kind of HFB
calculation is performed with the density-dependent
zero-range interaction obtained in a
previous study of the lead-mercury SD bands \cite{Ter95} (iv).

\section{The nucleus $^{150}$Gd}

On Figure \ref{Gdj2}, we have plotted the {\Gdz} dynamical moment
of inertia as a function of the rotational frequency $\ho$.
The HF calculation leads to an underestimation
of the moment of inertia at all spins.
In this region of high angular momentum,
the introduction of pairing correlations
induces an {\it increase} of the $\Jd$ moment of inertia which
allows to locally reproduce the data.
For instance, this is realized by the seniority force (iii)
for $\ho\approx$500keV. However,
the agreement is only obtained in the near vicinity of this angular frequency.
The first experimental point (see Fig. \ref{Gdj2}) is significantly
above the smooth trend of $\Jd$, which indicates a change in the structure
of the band.
This feature is not reproduced by this HFB calculation
(curve iii).
{}From the result (ii) one sees that the qualitative behavior
of $\Jd$ near $\ho=$400keV can be obtained by a further increase of the
seniority pairing strength. Unfortunately
this is obtained at the cost of an increase of $\Jd$
which  leads to a systematic overestimation.
It appears that the density dependent pairing interaction (iv)
leads to a much better global agreement: the HFB calculation
reproduces the rapid decrease
at low frequency while the magnitude
at large $\ho$ remains close to data.

Figure \ref{Gdep} shows the pairing energies
obtained in the three HFB calculations.
For the density-dependent interaction,
the  absolute value of the
pairing energies for neutrons are larger than those for protons.
With seniority forces we obtain the opposite result.
This can be understood as more neutrons orbitals are localized
in the vicinity of the nuclear surface than protons ones.
We note also that the two interactions (ii) and (iv), which lead
to a rapid decrease of the dynamical moment of inertia at low frequencies,
are also those which display the fastest decrease of the neutron
pairing energy at least up to $\ho\approx$0.55 MeV.
Beyond this value of the angular frequency the behaviour of
the pairing energy is much smoother.

Figures \ref{Gdern} and \ref{Gderp} give the particle routhians
for neutrons and protons respectively. The left-hand side of the figures
corresponds to the pure HF case. The right-hand side
shows the routhians defined as the eigenstates of the
cranked mean-field obtained with the density-dependent zero-range
interaction (iv).
The seniority
interactions (ii) and (iii) lead to very similar spectra.
The second N=7 neutron orbital is very close to the Fermi
level. It is fully occupied in the HF calculation.
On the other hand, the diagonal matrix element
of the density matrix in the basis associated with the cranked mean-field
is a rapidly varying function of $\ho$
when pairing correlations are included.
Above the Fermi level
these spectra show some significant differences
with those obtained with either a modified oscillator \cite{Fli95,Rag95}
or a Woods Saxon \cite {Naz89}.
For instance, the [512]3/2
and [523]5/2 neutron orbitals
are pushed up at higher excitation energies
while the [514]9/2 and the [402] 5/2 orbitals are very close.
On the other hand all calculations agree qualitatively
for orbitals below the Fermi level which are the one
related to SD bands in the lighter Gd isotopes.
The proton routhians are nearly identical in all calculations.
They differ only by the relative
positions of the [301]1/2 and [660]1/2 orbitals.
In our calculation, the first level is closer to the Fermi level, while
their order is reversed with both the modified oscillator and the Woods Saxon
potentials. The first orbitals above the Fermi level are
in good agreement.

Nazarewicz et al \cite{Naz89}
relate the low frequency behavior of the moment of inertia
to a crossing between the N=7 neutron quasiparticles around
$\ho=0.4$ MeV.
According to our quasiparticle (qp) plots
for the density-dependent
zero-range interaction (see Figure \ref{Gdeqp}),
an interaction between the [770]1/2
and [761]3/2 can also take place for $\ho$ just under
0.4 MeV. Unfortunately,
we could not extend the calculations
below the lowest frequency shown on Figure \ref{Gdeqp},
for numerical reasons. In any case the strength of this interaction
would be large, of the order of 0.65 MeV.
We find a similar value for
the energy difference between the [770]1/2 and the [761]3/2 qp's
in the HFB calculation with the seniority interaction
of the largest strength (ii). On the contrary with the
seniority interaction (iii) which predicts a smooth behavior of
$\Jd$ for low angular momentum frequencies, the
energy difference between the two qp's
is always larger than 1.6 MeV.
At high spins our calculation finds
a very gradual alignment of the N=7 qp orbitals.

As we have discussed, our HF and HFB calculations predict
rather different dependences  on $\ho$ of the occupation
of the [770]1/2 orbital in the canonical basis.
As deformation is mostly driven by this orbital,
an increase of pairing correlations which diminushes its
occupation affects the deformation of the nucleus.
This is reflected in
the evolution of the quadrupole moment
against rotational frequency (see Figure \ref{Gdqp}).
In the HF case, the quadrupole moment is steadily decreasing.
For the three HFB calculations, the deformation increases
very rapidly below a frequency
which depends on the intensity of pairing correlations.
Above this frequency, the quadrupole moment decreases with a slope
steeper than that found in the HF case.
The rise of $\Jd$ at low frequencies appears therefore related
to a complicated interplay between pairing correlations
and changes of deformation with rotation.
The value of the charge quadrupole moment that we obtain (16 eb) is
close to that measured for band 2 of $^{149}$Gd (15.6$\pm$0.3 eb)
which has the same number of intruder orbitals
as the yrast band of {\Gdz}\cite{Haa95}.
It is also significantly smaller than the value obtained
by the calculation of ref~\cite{Naz89} (16.9 eb).

Four excited SD bands have been observed in {\Gdz}.
The nature of their qp excitation content and in particular
the occupied intruder levels can be surmised from
the analogies between the behavior of their moment
of inertia with those of the yrast bands of neighbouring nuclei.
For band 2, the standard assignment of a N=6 and a [301]1/2 proton qp
excitations is compatible with our calculations.
Note however that the qp's corresponding to intruder states
are strongly mixed so that their hole
or particle nature is difficult to identify.
Band 3 and 4 \cite {Bea93}
are signature partners, with transition energies
close to 1/4 and 3/4 of the yrast band of {\Gdn}.
Therefore, the 2 qp excitations have been assigned
to a N=7 neutron orbital and either a [514]9/2 or a [402]5/2 excitation.
A possible interaction between band 4 and band 2 has made it
more plausible that bands 3 and 4 have negative parity.
Our qp diagrams are compatible with both assignements.
However in our calculations the [402]5/2 qp's interact with the
[651]1/2 (s=+1) or the [642]5/2 (s=-1) qp's around 0.55 MeV.
No sign of such an interaction is visible in the data.
Nevertheless, one must remember that the creation of qp's
may strongly modify the mean-field and shift the frequencies
at which interactions occur.
Band 5 \cite {Fal94} exhibits a pronounced discontinuity around
$\ho$=0.5 MeV. Above this frequency, the moment of inertia
presents similarities with that of
of {\Dyd}. It is therefore interpreted as a 4 proton
excitation  involving two N=6 and two [301]1/2
qp's. Our results are not incompatible with these
data. The HFB calculation using the surface peaked delta pairing
obtains indeed an interaction between two positive signature N=6
qp's at the correct energy. However, introducing such an excitation
into the HFB calculations would strongly modify the qp spectrum.
It is also probably as realistic to consider the {\Dyd} qp spectrum
(see Figure \ref{Dyeqp}) in which the excitation of the
[301]1/2 qp's is larger than 1.5 MeV. This large value does not
support the proposed scenario. However, because going from
{\Dyd} to {\Gdz} implies a lowering of the Fermi level by approximately
2.0 MeV, the [301]1/2 excitation may become more competitive.

\section{The nucleus $^{152}$Dy}
 The dynamical moment of inertia of {\Dyd} is plotted on
Figure \ref{Dyj2}. Calculations have been performed
for three of the four cases considered for {\Gdz}:
(i), (ii) and (iv). In the last two cases, as in all the results
reported in this work, the pairing strengths were kept equal to those
used in the study of {\Gdz}.
In all cases, the order of magnitude of the
moment of inertia is sligthly overestimated.
In the  HF calculation, both the neutron and proton contributions to the
moment of inertia are almost flat.
When pairing correlations
are included, the neutron contribution to $\Jd$ is always decreasing
with $\ho$, while the proton contribution is slightly
increasing up to 0.5 MeV. The nearly flat moment of inertia
results therefore from a quasi cancellation of both contributions.
Such a cancellation between the neutron and proton contributions to
$\Jd$ has also been obtained in ref~\cite{Sat94}, although in this case
the proton contribution to $\Jd$ is increasing faster
than in our calculations.
With the Skm$^*$ parametrisation
of the Skyrme force, it does not seem possible
to decrease the mean value of the
moment of inertia of {\Dyd} and thus obtain a
better agreement with data by
modifying only the pairing interaction. The HF moment of inertia
shows the same quality of agreement with data as obtained
by Satu{\l}a~\cite{Sat94} and as the relativistic
Hartree calculation of Konig and Ring \cite{Kon93}.
The calculation of Shimizu et al \cite{Shi90}
leads also to an overestimation of the dynamical moment
of inertia, probably related to a deficiency of the
Nilsson potential parametrization.
In our HFB calculations we find that the pairing energies are always non zero,
although their smooth variations with angular frequency lead only
to small contributions to the moment of inertia.

The variation of the charge quadrupole deformations
as a function of the rotational frequency for the three
cases shown on Figure \ref{Dyj2} is plotted on Figure \ref{Dyqp}.
As for {\Gdz}, the HF quadrupole moment is always decreasing, while the
pairing correlations lead to an increase of this moment.
However, the effect is less pronounced for Dy than for Gd.
The mean deformation that we obtain (17.5 eb) is very close
to that measured for band 4 of $^{149}$Gd
and for the lowest band of {\Dyd}\cite{Haa95}.
It is lower than the value obtained
in of ref~\cite{Naz89} (18.9 eb).

On Figures \ref{Dyern}, \ref{Dyerp} and  \ref{Dyeqp}
are plotted the energies of the neutron and proton routhians and
of the quasiparticules respectively, as a function of the rotational
frequency. The differences between the Gd and Dy neutron routhians
are larger in the HF than in the HFB calculations.
The main difference is the slightly deeper position of the
[770]1/2 levels with respect to the Fermi level, so that the corresponding
qp has always a positive energy. This prevents the crossing
between the two N=7 qp's. Our neutron diagrams differ from the
cranked Woods-Saxon ones by the position of the [411]1/2 levels,
which are lower and more separated from the N=6 levels in our calculation
than in ref \cite{Naz89}. Similarly we find a [301]1/2 proton routhian
more bound and not as close to the intruder N=6 levels.

Our particle and qp diagrams are compatible with the experimental
data on excited bands in {\Dyd}\cite{Dag94} for bands based
on 2qp neutron excitations (band 4, 5 and 6). Our results also
support the contention that band 3 could be constructed on the
excitation of the [651]3/2 and [530]1/2 proton qp's.
At the deformation of the yrast SD bands, the [301]1/2 proton qp is
very excited. Because we note that
at larger deformations this routhian
becomes closer to the Fermi level we expect an energetically competitive
excited band based on this qp to be more deformed.
A simultaneous excitation of the rapidly
downsloping and deformation driving N=7
proton orbital could realize this effect. Therefore the
assignment of such a qp excitation content to experimental band 2
is plausible according to our spectrum. Note that our
N=7 proton particle state is more excited than in
calculations based on Saxon-Woods potentials and is always
above the [530]1/2 qp routhian of negative signature.
This feature is in agreement with the larger intensity
observed for band 3 than for band 2.

\section{The nucleus $^{151}$Dy}
The neutron spectrum of {\Dyd} (Figure \ref{Dyern}) indicates that
the ground SD band of {\Dyu} should correspond to
a hole in the [770]1/2 intruder state.
Excited bands can be
obtained by making another hole in one of the neutron states
located 1.0 MeV below this intruder. The spectra
displayed on  Figure \ref{Dyern} are compatible
with the assignements of two of the four excited bands observed
experimentally \cite{Nis95} to the neutron
[642]5/2 (band 2) and [411]1/2 (band 4) orbitals. The
large splitting between the two [411]1/2 levels is also
in agreement with the fact that the
signature partner of band 4 is not observed. The other bands
should correspond to the  [651]1/2 orbitals.
Based on these excitations on can construct
up to three bands of the same parity and of both
signatures with orbitals which are strongly mixed
by rotation. The study of these excited bands is rather
delicate and will not be performed here.

The HF moment of inertia of the yrast SD band of {\Dyu} is
displayed on Figure \ref{Dyuj2}, it shows a smooth decrease as a function
of $\ho$ which is slightly more pronounced than that found for {\Dyd}.
This feature that our calculation appears to share with all
other self-consistent calculations is in contradiction
with the behavior which could be expected based
on the qualitative discussion of ref.\cite{Ben88}.
It disagrees also with data which shows a slow regular increase
of $\Jd$ over the entire range of observed angular momenta.
It has been suggested that a gradual mixing with another configuration
might resolve the discrepancy between calculation and data \cite{Naz89}.
In principle pairing correlations can introduce such a mixing.
A HFB calculation with the seniority force (ii) shows indeed
a sensitivity of $\Jd$ to pairing. On the other hand as in ref.\cite{Naz89}
we find a curve whith a maximum near $\ho=$500keV.
An analysis of the contributions to $\Jd$ shows that the
neutron contribution is almost flat (instead of decreasing in
{\Dyd} in agreement with the
discussion of ref.\cite{Ben88}). The maximum is therefore due to
the proton contribution.
Finally we note that the charge quadrupole moment predicted
by our HFB calculation is of 17 eb, a value smaller than the 18 eb
obtained in ref~\cite{Naz89}.

\section{The nucleus $^{151}$Tb}

Although many bands have been detected in this nucleus
we have limited our calculations to the first three bands
which will allow a test of our proton spectrum.
Since pairing correlations are not expected to play
a major role, we have only performed a HF study.

{}From the consideration of the proton routhian spectrum of {\Gdz}
(Figure \ref{Gderp})
it appears favorable to define bands in {\Tbu} by the creation
of qp based on the [651]3/2 orbitals. On the other hand,
the qp routhian spectrum of {\Gdz} suggests that at high spins,
one considers also the qp associated with the orbital [301]s=$-$.
{}Finally a fourth band could be build on the [301]s=+ qp.
In fact four of the eight bands that have been observed
\cite {Def94,Kha95} fit rather well this
assignment pattern. The
excitations which have been used to construct the three bands
studied in this work are presented on figure \ref{Tberp}.
This figure shows the effect that each breaking of a pair
generates in the proton routhian spectrum.
In the
limit of small angular frequency, the time-reversal-symmetry breaking
always lead to a situation in which the empty orbit is more bound than
its occupied signature partner. Apart from the two routhians of
the broken pair, the rest of the spectrum is not very much modified.

The moments of inertia obtained for all these
configurations are shown on the right-hand
 side of Figure \ref{Tbuj2} together with
the experimental data (left). The overall agreement with data is quite good.
Experimentally, band 2  which is contructed on the [301]1/2 s=+ orbital
and has therefore the same proton intruder content as {\Dyd}
is identical to the yrast band of this nucleus.
Our calculation finds indeed that the $\Jd$ curves of these two bands are
almost the same.
This confirms the special nature of the [301]1/2 orbital.
It shows that mean-field calculations are relevant for the problem
of identical bands whether high-$K$
(see ref.\cite{Che92}) or low-$K$ (this work)
are involved. However, for the [301]1/2 orbital,
our calculation does not lead to identical bands
because we do not predict the correct relative alignment. A discussion of
this effect and the first indications on how it may be related to the
time-odd components of the mean-field can be found in ref.\cite{Dob95}.
The data on the relative magnitudes of the three
moment of inertia are very well reproduced.
Our calculations predict close values for the charge quadrupole moments of
band 2 and of the identical yrast band of {\Dyd}. On the other hand
the values of these moments for bands 1 and 3 are 1 eb lower.
At $\ho$ around 0.7 MeV,
the excitation energies of bands 2 and 3 with respect  to
band 1 are of the order of 1.1 MeV and 0.8 MeV respectively.
This is in good agreement with the feeding pattern of band 3,
but is probably too large for band 2.

\section{Conclusion}

 In this work, we have studied the lowest superdeformed bands
of four nuclei in the A=150 mass region. We have compared
data with results obtained by the cranked Hartree-Fock method on the one hand
and with the cranked Hartree-Fock-Bogoliubov method on the
other hand. In the latter case, we have tested three different forces
in the pairing channel.

Our calculations reproduce well the magnitude of moments of inertia.
With the notable exception of {\Dyu} they also describe correctly
the evolution of $\Jd$ versus angular momentum.
Our results confirm also the general belief that pairing correlations
are much less important in this mass region than for Hg and Pb isotopes.
Only for {\Gdz}
do we find that pairing  correlations are crucial to describe the
behaviour of its moment of inertia
for angular momenta  between 30 and 40$\hbar$.
We nevertheless find that even
at the highest observed spins
where pairing correlations are much weakened, they still affect
the magnitude of the $\Jd$.
We also checked that
the moment of inertia of {\Gdz} depends sensitively on the features
of the pairing force.
In particular the steeply down going slope of $\Jd$
is correctly reproduced with a zero-range density-dependent
while a very poor agreement with experiment is obtained
with seniority (monopole) interactions.

Our discussion of excited bands in Gd and Dy isotopes shows that
a Skyrme force like Skm$^*$ leads to a correct description of
the properties of the high spin behavior of the mean-field.
The calculated charge quadrupole moments of the
SD bands are very close
to the latest Eurogam values \cite{Haa95}.
We believe however that this agreement is
especially significant for the variation of the quadrupole
moments between SD bands of {\Gdz} and {\Dyd} since
the experimental uncertainties on the absolute values remain large.
The lowest qp excitations that we predict
are compatible with the characteristics
of the excited bands found in nuclei of this mass region.
In $^{151}$Tb, a calculation  of the three lowest bands
leads to a correct prediction of
the relative values of the moments of inertia within this nucleus
and also with neighbboring isotopes.

Following encouraging results obtained in the A=190 mass region
on the identity of the rotational bands
associated with specific quasi-particle content of the mean-field\cite{Che92},
the present work reproduces also the identity of the ground state
band of {\Dyd} and of
band 2 of {\Tbu}. This gives us some confidence
on the ability of mean-field studies at dealing with the
identical-band question.
On the other hand, a recent study \cite{Dob95} has shown
that a detailed understanding may require a thorough
investigation of the time-odd terms in
the Skyrme functional.

A remarkable outcome of this study is that
the parametrisation of the density-dependent zero-range interaction,
adjusted on the dynamical moment of inertia of {\Pbq} at low spins,
without any modification,
leads to very satisfactory results in the A=150 mass region
and at higher spins.

This parametrisation shown to be successful on the neutron-poor
side of the stability valley remains to
be tested in cases that are more sensitive to the isospin
degree of freedom.
Although
we have chosen a force with identical
neutrons and protons pairing strengths,
we believe that a relative variation of
a few percents would not significantly
modify our predictions concerning
the moments of inertia of SD bands of the nuclei
studied in this work.
The study of other regions of the
mass table is thus necessary to better determine the isospin behavior
of the pairing force.
The present work can be extented in several other directions.
For instance, we still have
to test our pairing interaction
on the properties of rotational bands at normal deformations
and at low spins.
In particular,  it is important
to address the question of low-spin identical bands as it should
be more sensitive to the interplay between the mean-field and the pairing
channels.
In addition the study of a carefully selected subset of the
observed excited SD bands should
also provide a demanding test of
the ability of our zero-range density-dependent pairing
interaction to describe bands with various quasi-particle excitations
leading to very different dependence of $\Jd$ on the angular frequency.
Work along these directions is in progress.

\noindent{\bf Acknowledgements}

This work was supported in part by the contract PAI-P3-043
of the Belgian Office for Scientific Policy.
We thank J. Dobaczewski, B. Haas, R. Janssens, W. Nazarewicz, I. Ragnarsson,
and R. Wyss, for discussions and useful comments and J.-P. Blaizot
for a critical reading of the manuscript.

\newpage
\begin{center}
\bf Figures
\end{center}

\begin{figure}[ht]
\caption {
Comparison of the experimental dynamical moment of inertia of {\Gdz} (dots)
as a function of angular velocity with
the results of four mean-field calculations:
pure HF  (full line), HFB with a
seniority pairing adjusted to  the S$_{2n}$ (dashed line),
with a reduced seniority pairing (see text)
(dash-dotted line),
and with a surface-active delta pairing (dotted line).
}
\label{Gdj2}
\end{figure}

\begin{figure}[ht]
\caption[F2]{
Neutron and proton pairing energies
of the {\Gdz} ground SD band
obtained with different pairing interactions:
two  seniority pairing interactions (see text),
case (ii) - dashed line-, case (iii) - dash-dotted line-,
zero-range  interaction
- dotted line-.
}
\label{Gdep}
\end{figure}

\begin{figure}[ht]
\caption[F3]{
Neutron particle routhians of {\Gdz} calculated
without pairing (left) and with
the zero-range
 pairing interaction (right). The (parity, signature)
combinations
are indicated by a full line (+,+), a dashed line
(+,$-$), a dot-dashed line ($-$,+)
and a dotted line ($-$,$-$).
}
\label{Gdern}
\end{figure}

\begin{figure}[ht]
\caption[F4]{
Proton particle routhians of {\Gdz} calculated
without pairing (left) and with
the zero-range
pairing interaction (right). The (parity, signature)
combinations
are indicated by a full line (+,+), a dashed line
(+,$-$), a dot-dashed line ($-$,+)
and a dotted line ($-$,$-$).
}
\label{Gderp}
\end{figure}

\begin{figure}[ht]
\caption[F5]{
Neutron (left) and proton (right)
quasi particle routhians of {\Gdz} calculated
with the zero-range
pairing interaction. The (parity, signature)
combinations
are indicated by a full line (+,+), a dashed line
(+,$-$), a dot-dashed line ($-$,+)
and a dotted line ($-$,$-$).
}
\label{Gdeqp}
\end{figure}

\begin{figure}[ht]
\caption[F6]{
Charge quadrupole moments of the SD band of {\Gdz}
calculated in the HF case (full line),
with the seniority pairing interactions
(dashed line and dashed-dotted lines),
andd with the surface-active delta pairing (dotted line).}
\label{Gdqp}
\end{figure}

\begin{figure}[ht]
\caption[F7]{
Comparison of
the experimental dynamical moment of inertia of
{\Dyd} (dots)
as a function of angular velocity with
the results of calculations
without pairing (full line)
with a seniority pairing pairing (dashed line)
and surface-active delta pairing (dotted line).
}
\label{Dyj2}
\end{figure}

\begin{figure}[ht]
\caption[F9]{
Charge quadrupole moments of the {\Dyd} SD band
calculated in the HF case (dotted line),
with a seniority pairing interaction
(dashed line),
and with the surface-active  delta pairing (dotted line).}
\label{Dyqp}
\end{figure}

\begin{figure}[ht]
\caption[F10]{
Neutron particle routhians of {\Dyd} calculated
without pairing (left) and with
the surface-active
(right) pairing interaction. The (parity, signature)
combinations
are indicated by a full line (+,+), a dashed line
(+,$-$), a dot-dashed line ($-$,+)
and a dotted line ($-$,$-$).
}
\label{Dyern}
\end{figure}

\begin{figure}[ht]
\caption[F11]{
Proton particle routhians of {\Dyd} calculated
without pairing (left) and with
the surface-active
(right) pairing interaction. The (parity, signature)
combinations
are indicated by a full line (+,+), a dashed line
(+,$-$), a dot-dashed line ($-$,+)
and a dotted line ($-$,$-$).
}
\label{Dyerp}
\end{figure}

\begin{figure}[ht]
\caption[F12]{
Neutron (left) and proton (right)
quasi particle routhians of {\Dyd} calculated
with the surface-active
pairing interaction. The (parity, signature)
combinations
are indicated by a full line (+,+), a dashed line
(+,$-$), a dot-dashed line ($-$,+)
and a dotted line ($-$,$-$).
}
\label{Dyeqp}
\end{figure}

\begin{figure}[ht]
\caption[F13]{
Comparison of the {\Dyu} experimental dynamical moment of inertia
(open squares)
as a function of  angular velocity with
the results of a pure HF calculation (full line)
and a HFB calculation (case ii).
}
\label{Dyuj2}
\end{figure}

\begin{figure}[ht]
\caption[F14]{
Proton particle routhians of {\Tbu} calculated
without pairing for three different HF configurations.
The filled and empty orbitals near the Fermi level
are marked with filled and open circles respectively.
 The (parity, signature)
combinations
are indicated by a full line (+,+), a dashed line
(+,$-$), a dot-dashed line ($-$,+)
and a dotted line ($-$,$-$).
}
\label{Tberp}
\end{figure}

\begin{figure}[ht]
\caption[F15]{
Dynamical moment of inertia of
three bands in {\Tbu}
as a function of angular velocity.
Left: experiment, right: HF calculation. For comparison,
the $\Jd$ of {\Dyd} ground SD band is also shown.
}
\label{Tbuj2}
\end{figure}


\newpage
\begin{thebibliography}{99}

\bibitem{Ben88}
               T. Bengtsson, I Ragnarsson and S. \AA berg,
               Phys. Lett. {\bf B208} (1988) 39

\bibitem{Naz89}
                W. Nazarewicz, R. Wyss and A. Johnson,
                Phys Lett. {\bf B225} (1989) 208

                W. Nazarewicz, R. Wyss and A. Johnson,
                Nucl. Phys. {\bf A503} (1989) 285

\bibitem{Sat94}
               W. Satu{\l}a, R. Wyss and P. Magierski,
               Nucl. Phys.  {\bf A578} (1994) 45

\bibitem{Shi90}
               Y.R. Shimizu, E. Vigezzi and R.A. Broglia,
               Nucl. Phys.  {\bf A509} (1990) 80


\bibitem{Gal94}
               B. Gall, P. Bonche, J. Dobaczewski,
               H. Flocard and P.-H. Heenen,
               Z. Phys. {\bf A348} (1994) 183

\bibitem{Hee95}
               P.-H. Heenen, P. Bonche and H. Flocard,
               Nucl. Phys. {\bf A588} (1995) 490


\bibitem{Ter95}
               J. Terasaki, P.-H. Heenen, P. Bonche, J. Dobaczewski,
               and H. Flocard,
               Nucl. Phys. in press
\bibitem{Haa95}
               B. Haas, private communication.



\bibitem{Fli95}
               S. Flibotte et al.,
               Nucl. Phys. {\bf A584} (1995) 373

\bibitem{Rag95}
               I. Ragnarsson, private communication.

\bibitem{Bea93}
               C. Beausang et al.,
               Phys. Rev. Lett. {\bf 71} (1993) 1800

\bibitem{Fal94}
               P. Fallon et al.,
               Phys. Rev. Lett. {\bf 73} (1994) 782
\bibitem{Kon93}
               J. Konig and P. Ring,
               Phys. Rev. Lett. {\bf 71} (1993) 3079
\bibitem{Dag94}
               P.J. Dagnall et al.,
               Phys. Lett. {\bf B335} (1994) 313

\bibitem{Nis95}
               D. Nisius et al.,
               Phys. Lett. {\bf B346} (1995) 15

\bibitem{Def94}
               G. de France et al.,
               Phys. Lett. {\bf B331} (1994) 290

\bibitem{Kha95}
               B. Kharraja et al.,
               Phys. Lett. {\bf B341} (1995) 278

\bibitem{Che92}
               B.Q. Chen, P.-H. Heenen, P. Bonche, H. Flocard and M.S. Weiss,
               Phys. Rev. {\bf C46} (1992) 1582

\bibitem{Dob95}
               J. Dobaczewski and J. Dudek,
               preprint CRN 95-11

%
%
%



\
\end{thebibliography}
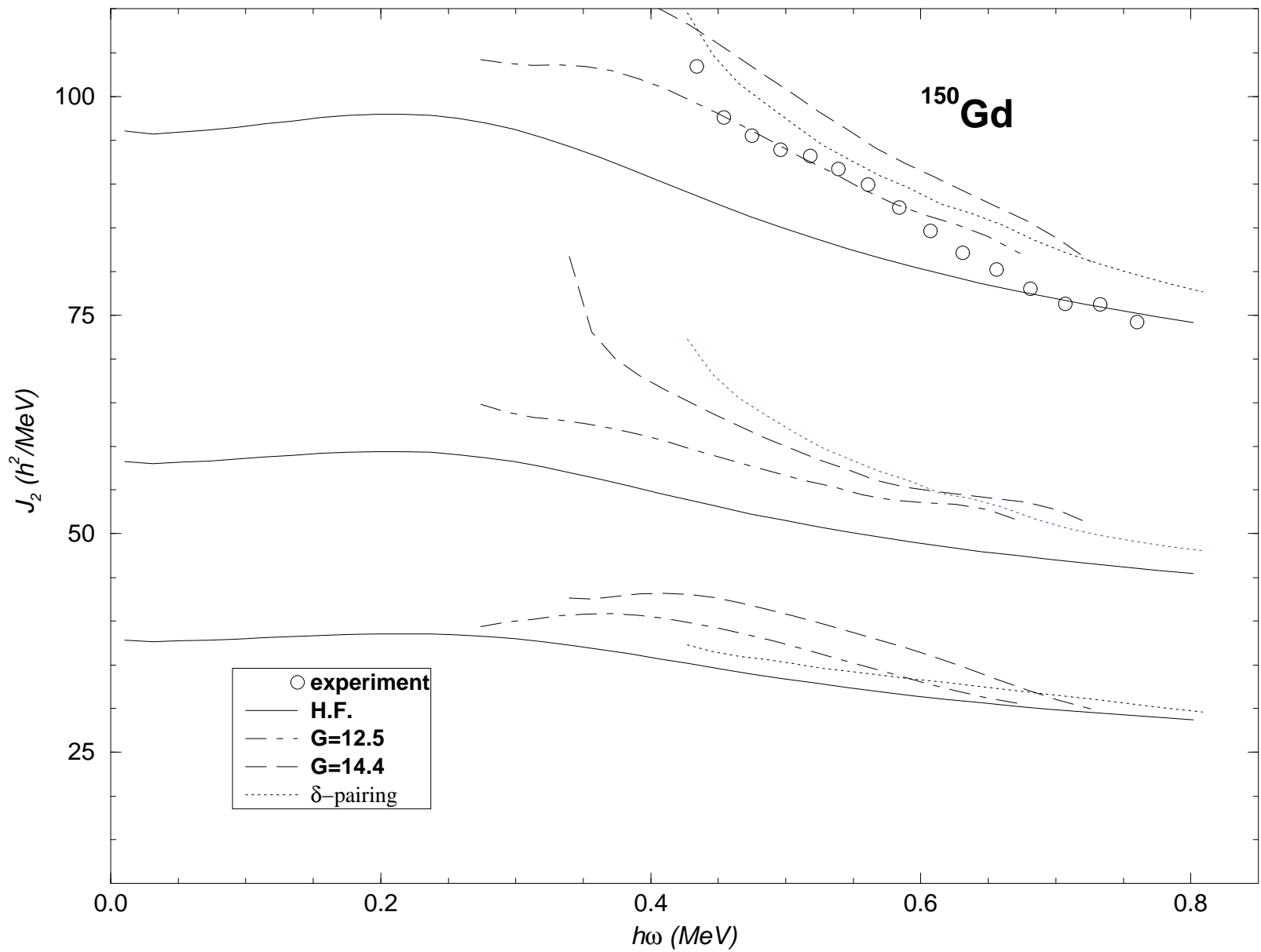
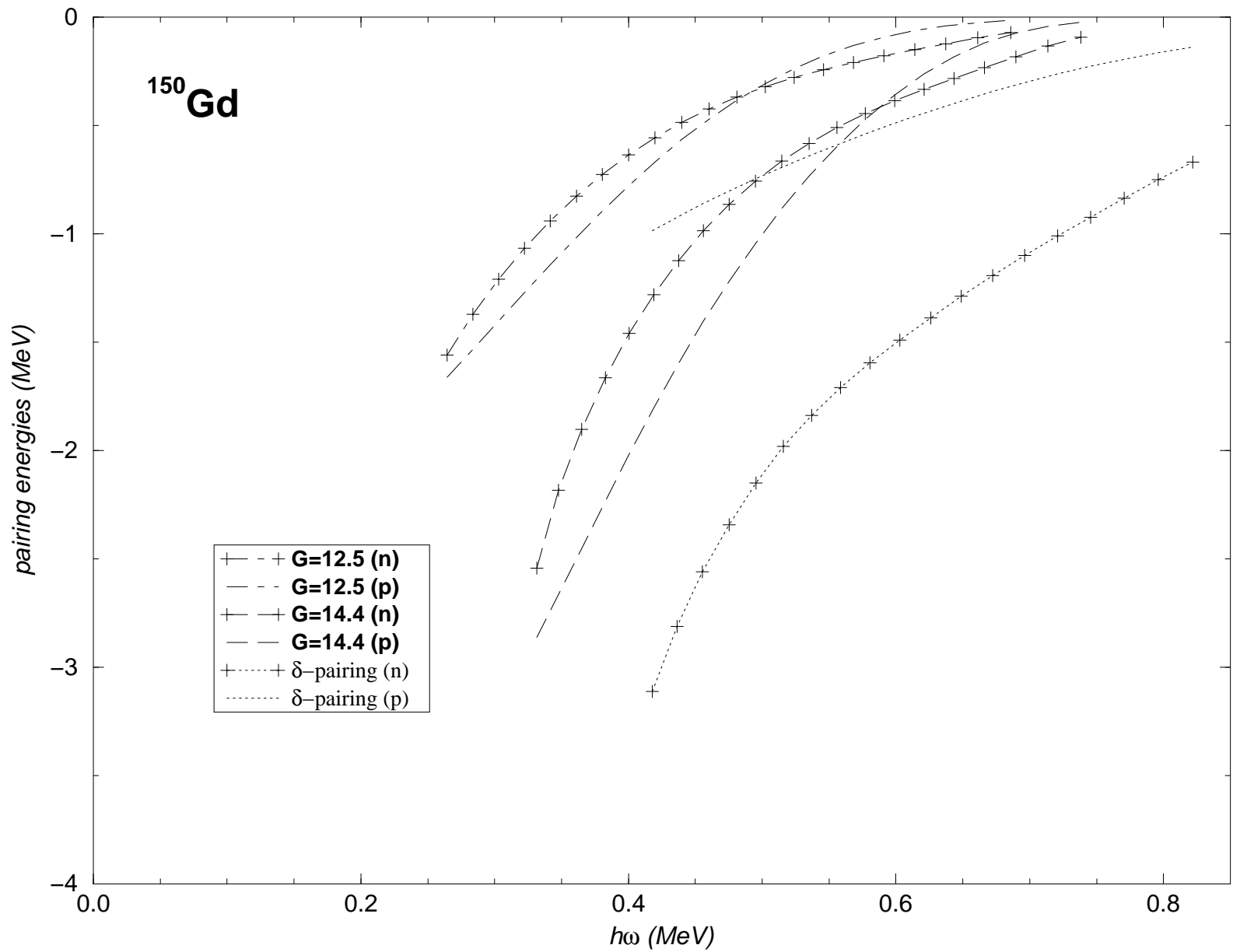
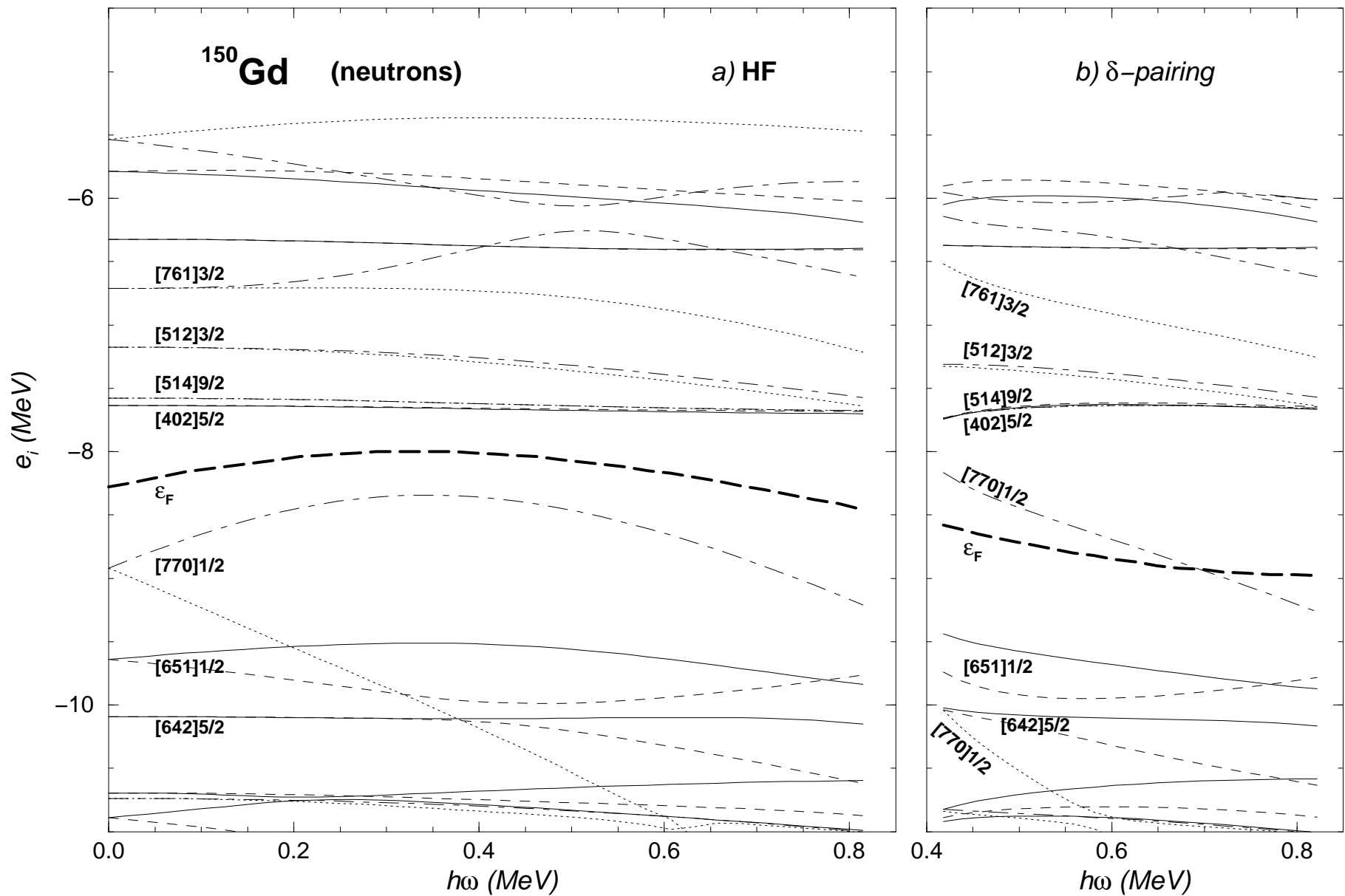
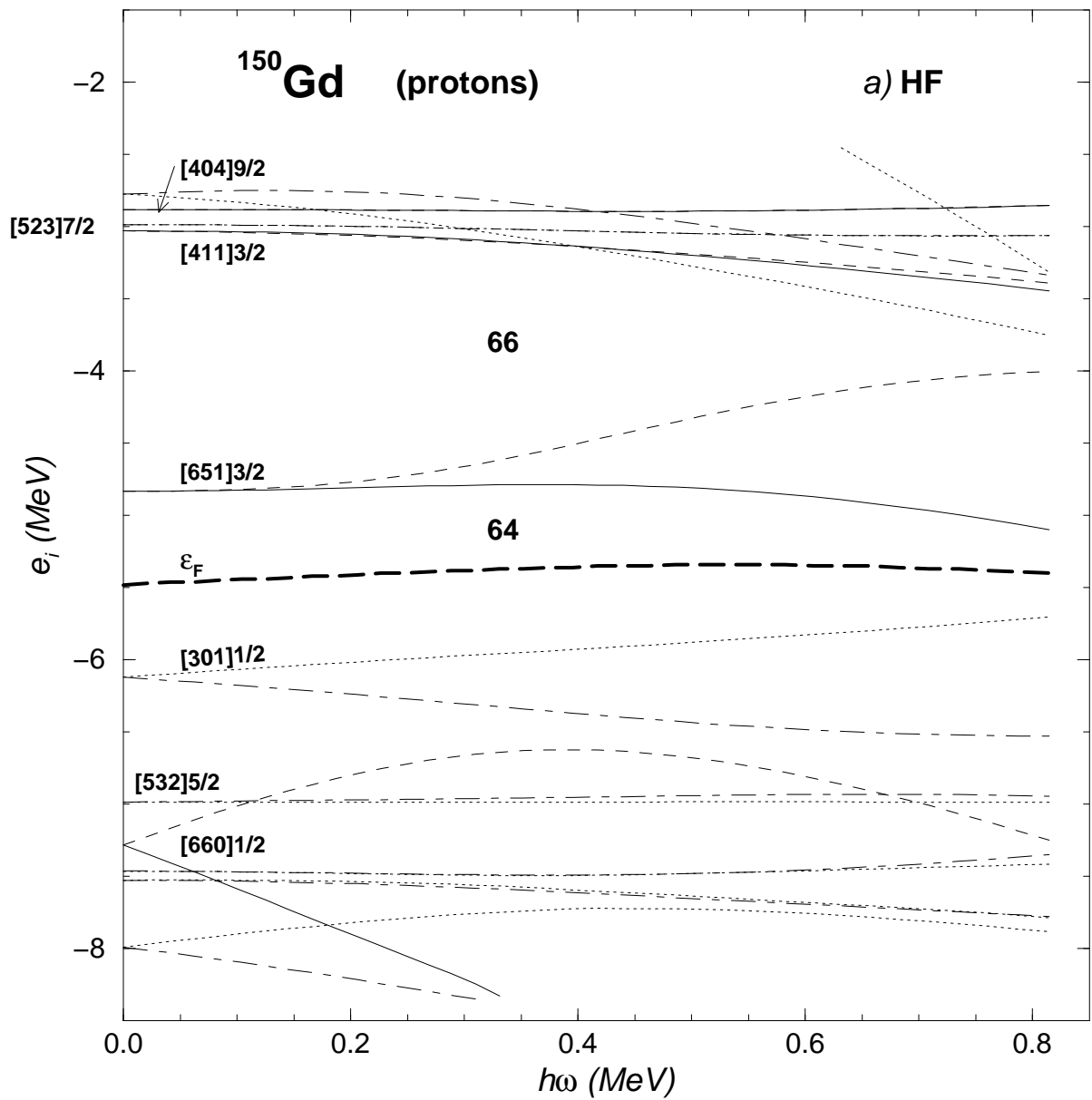
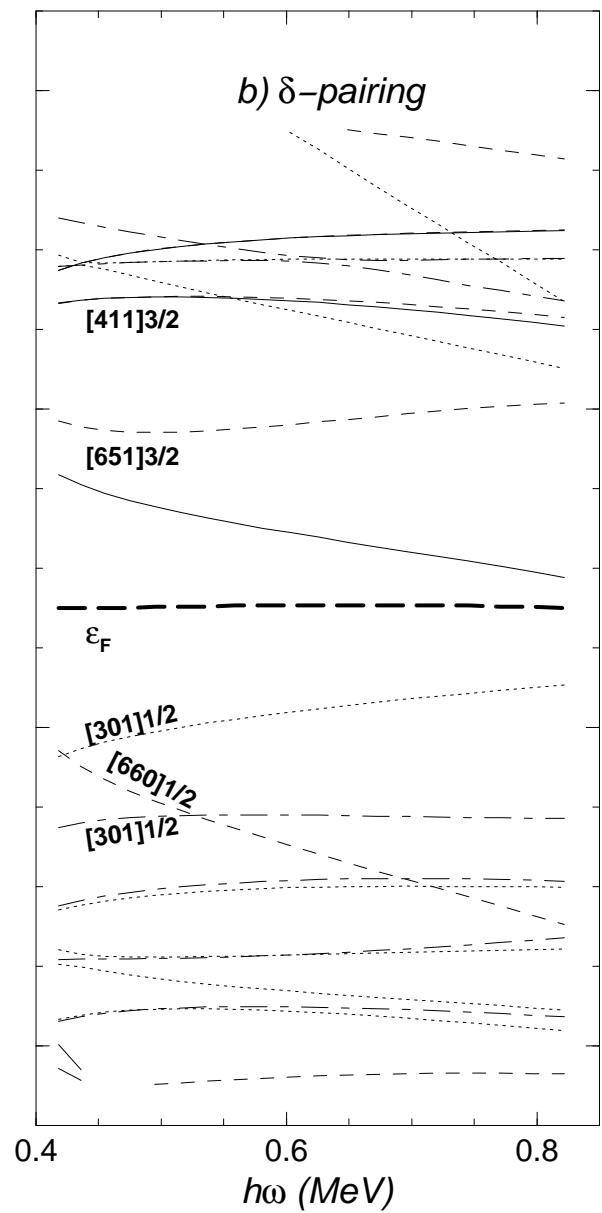
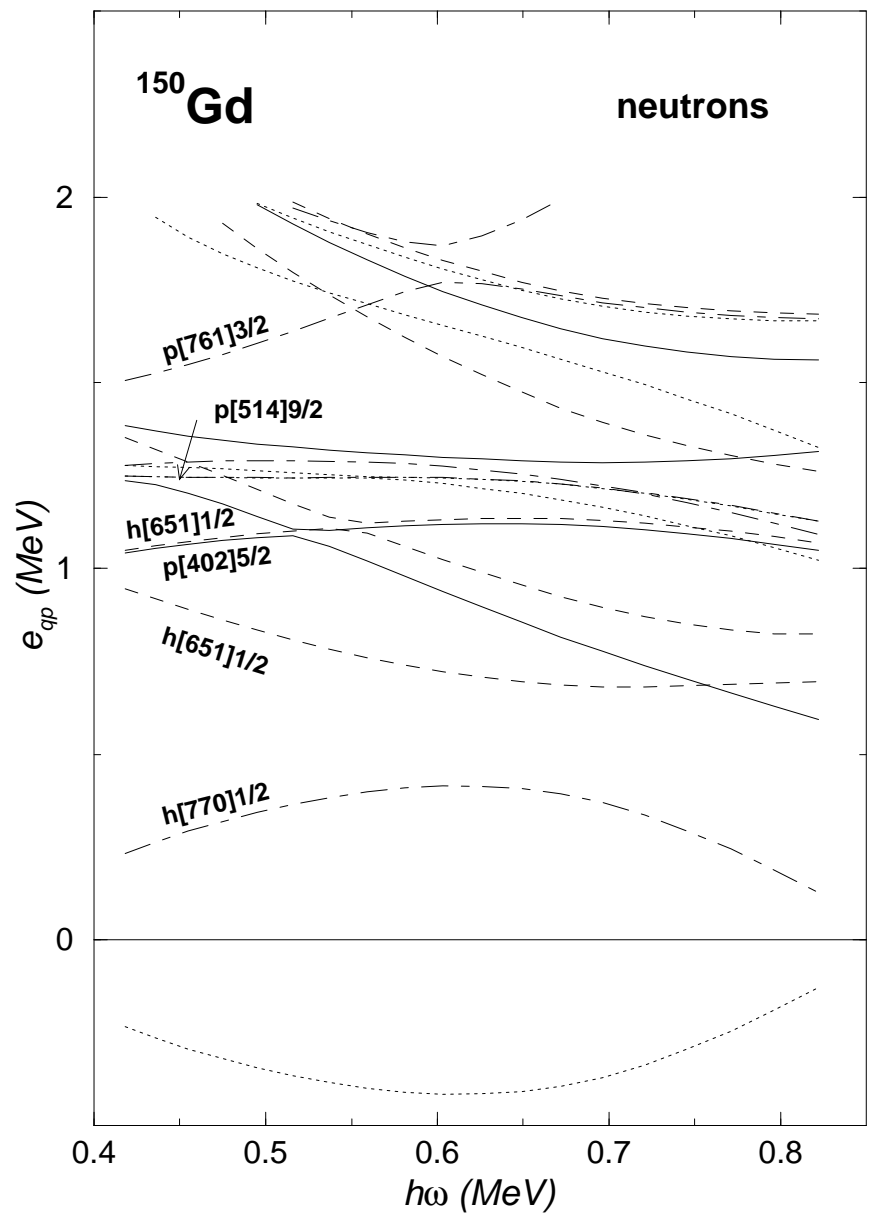
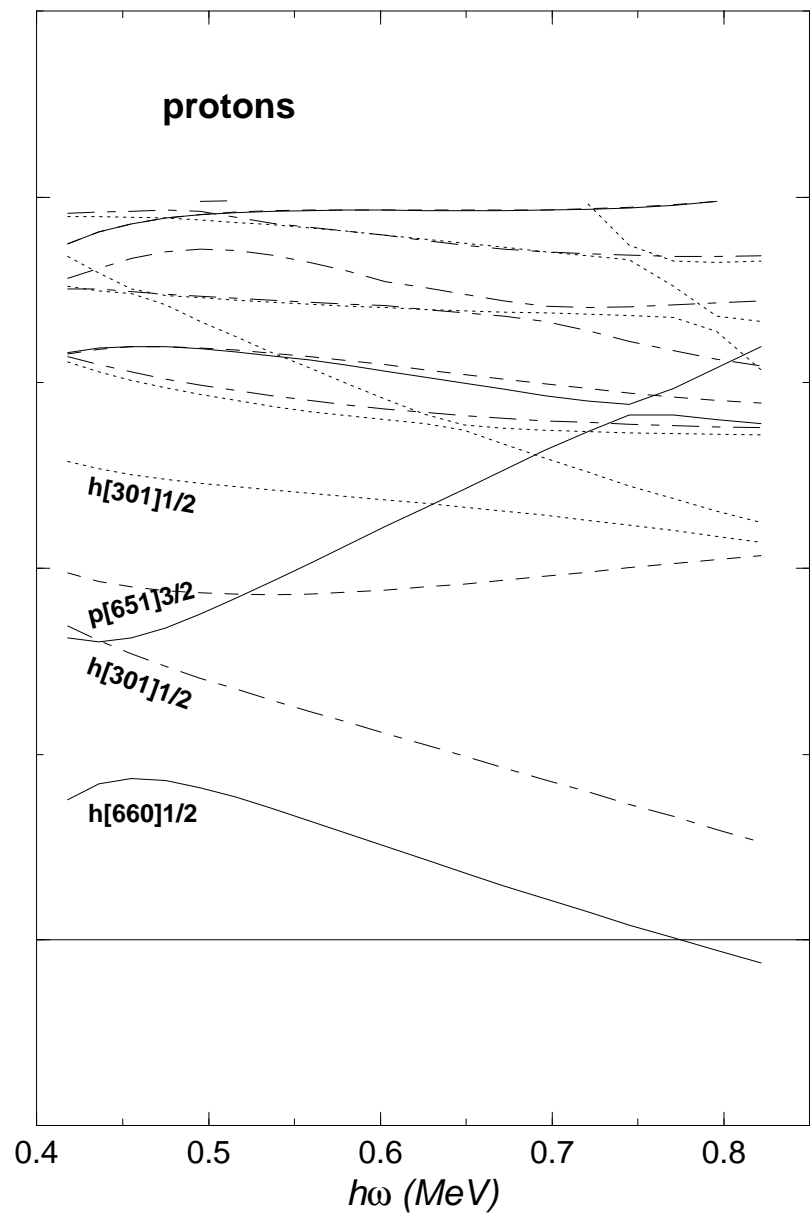
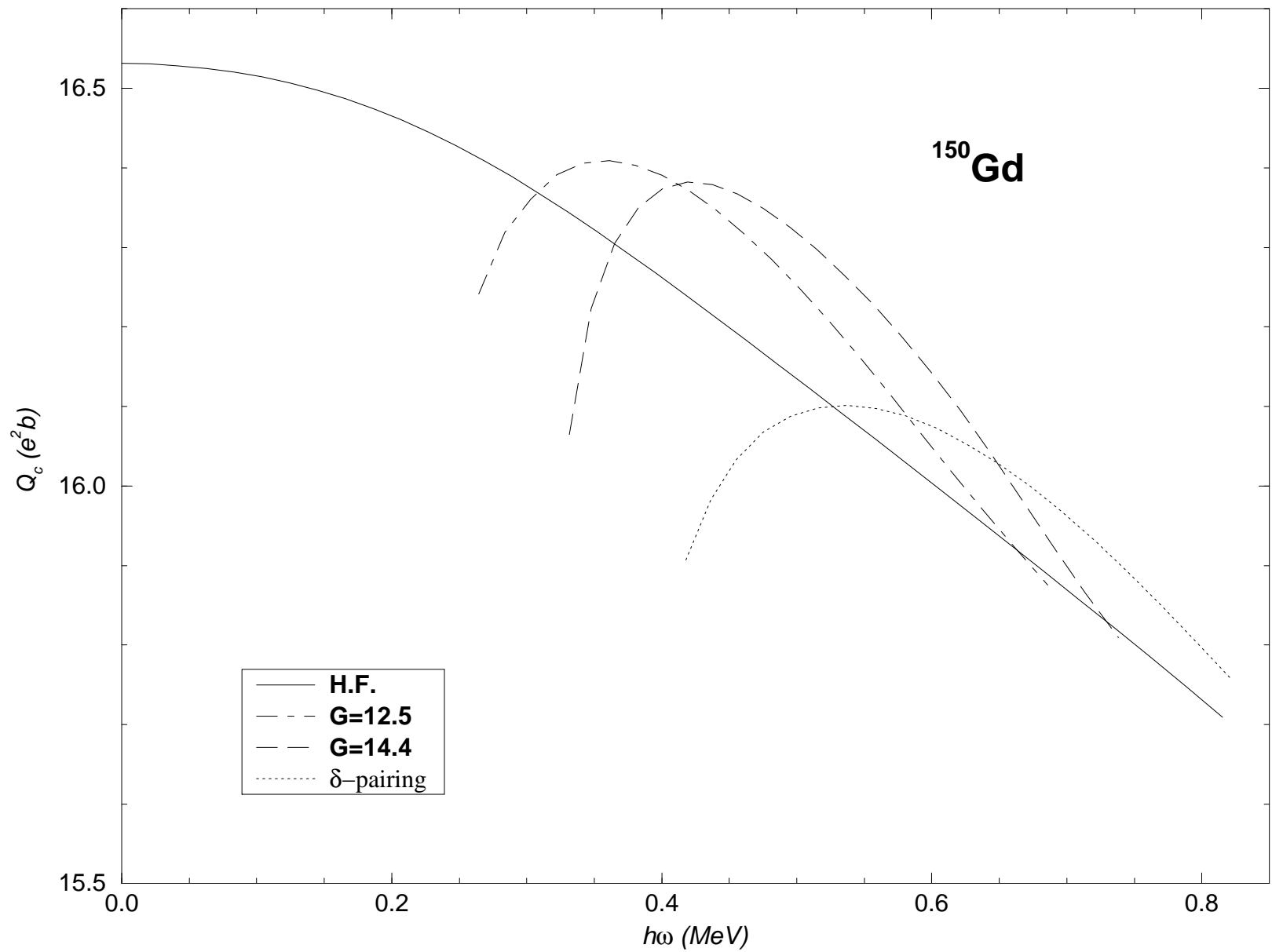
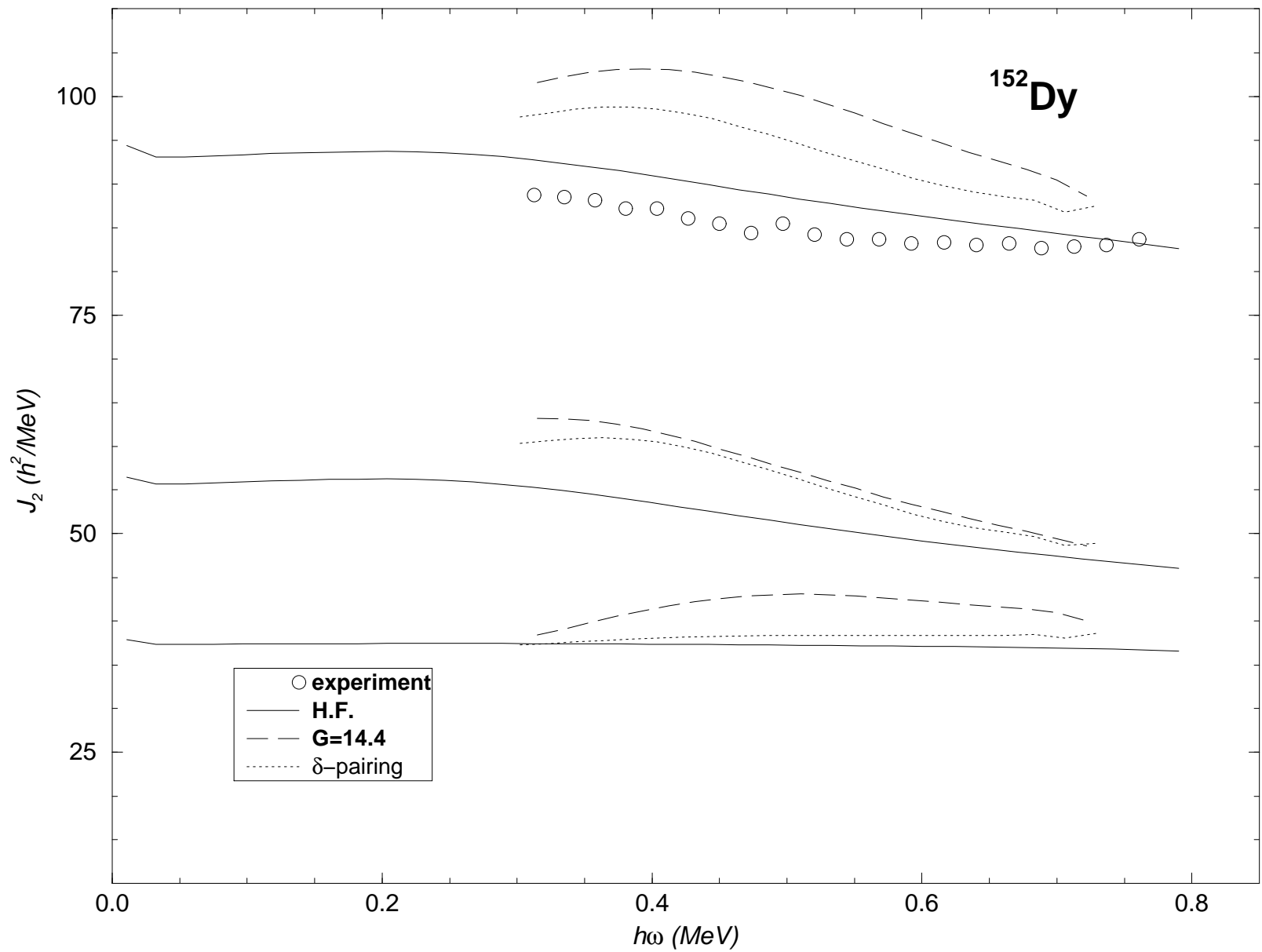
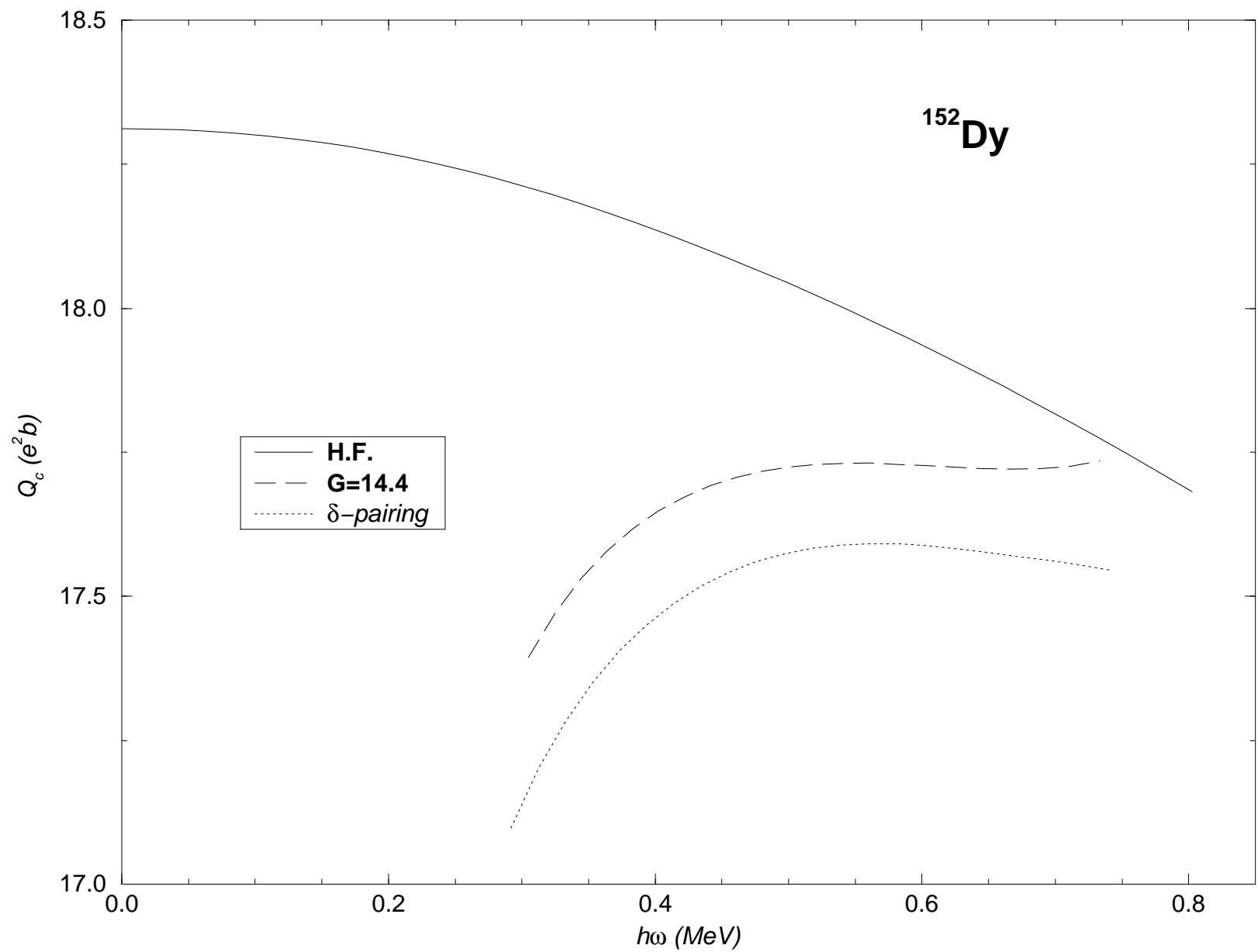
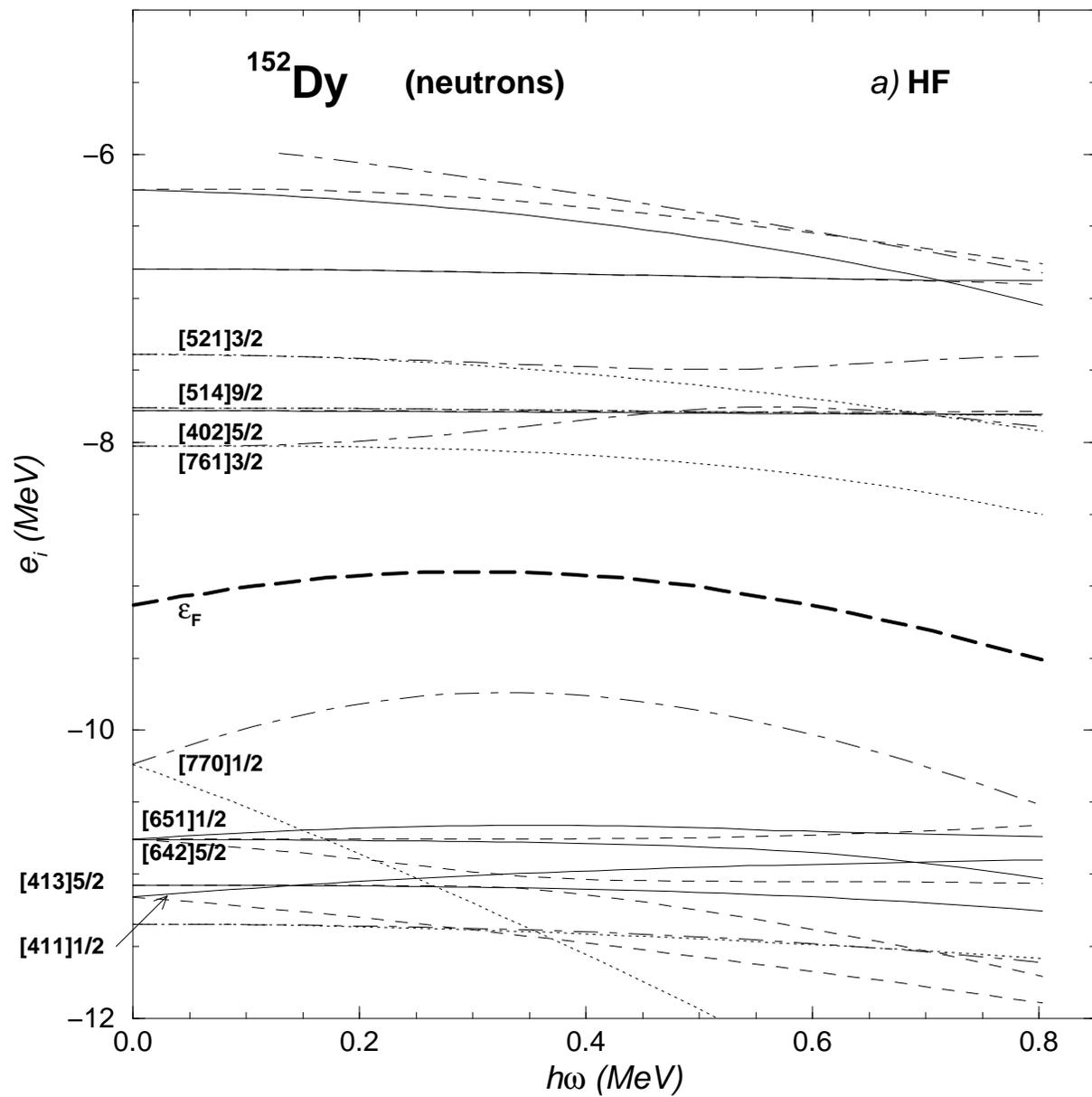
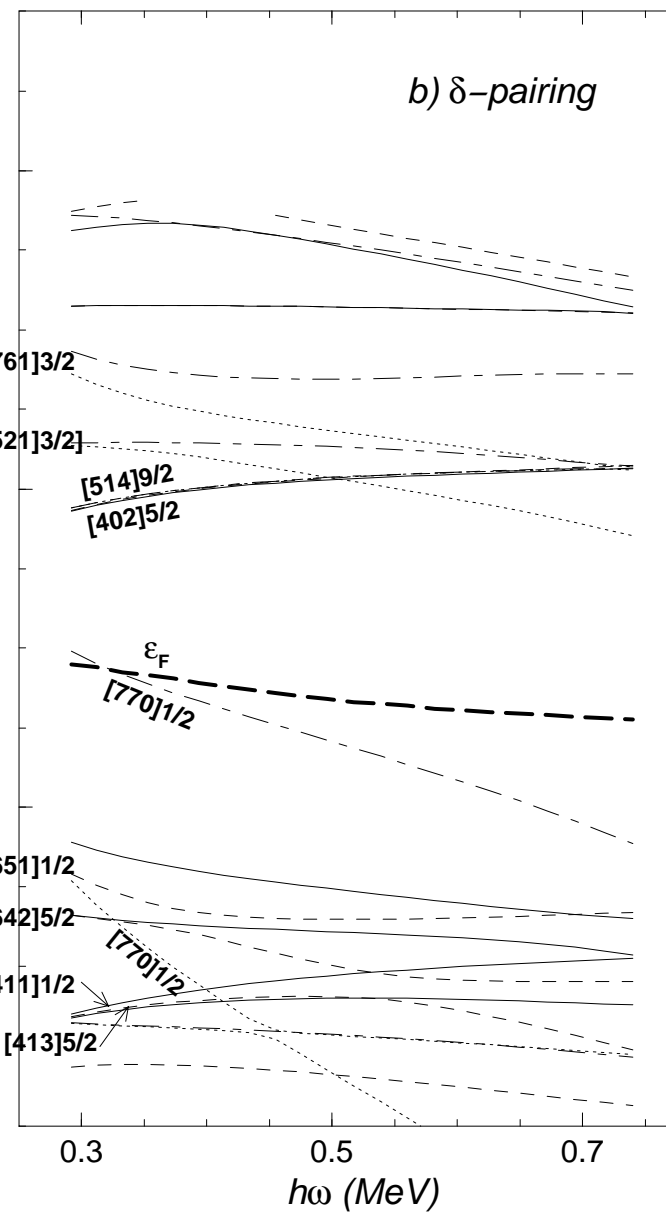
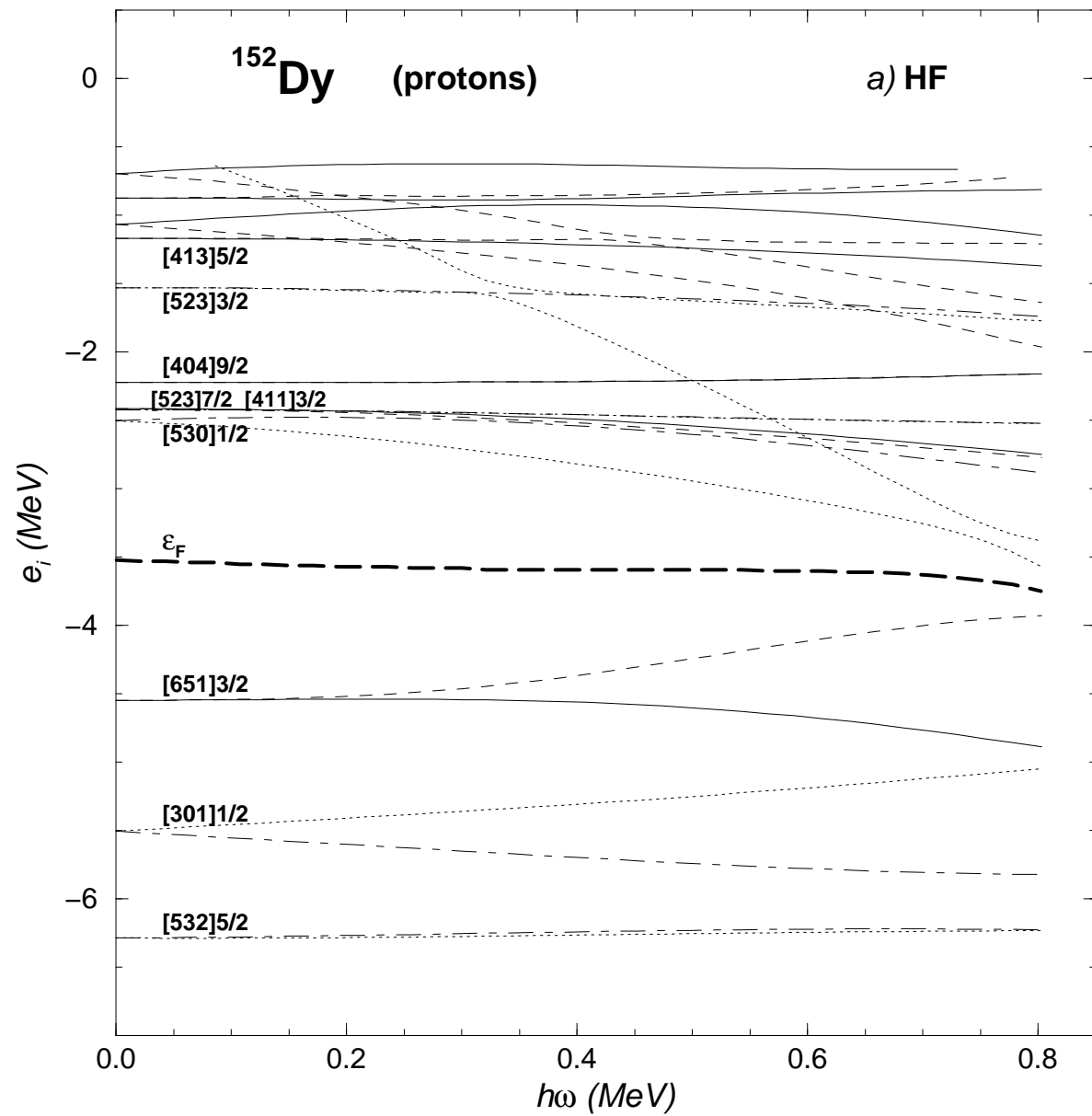
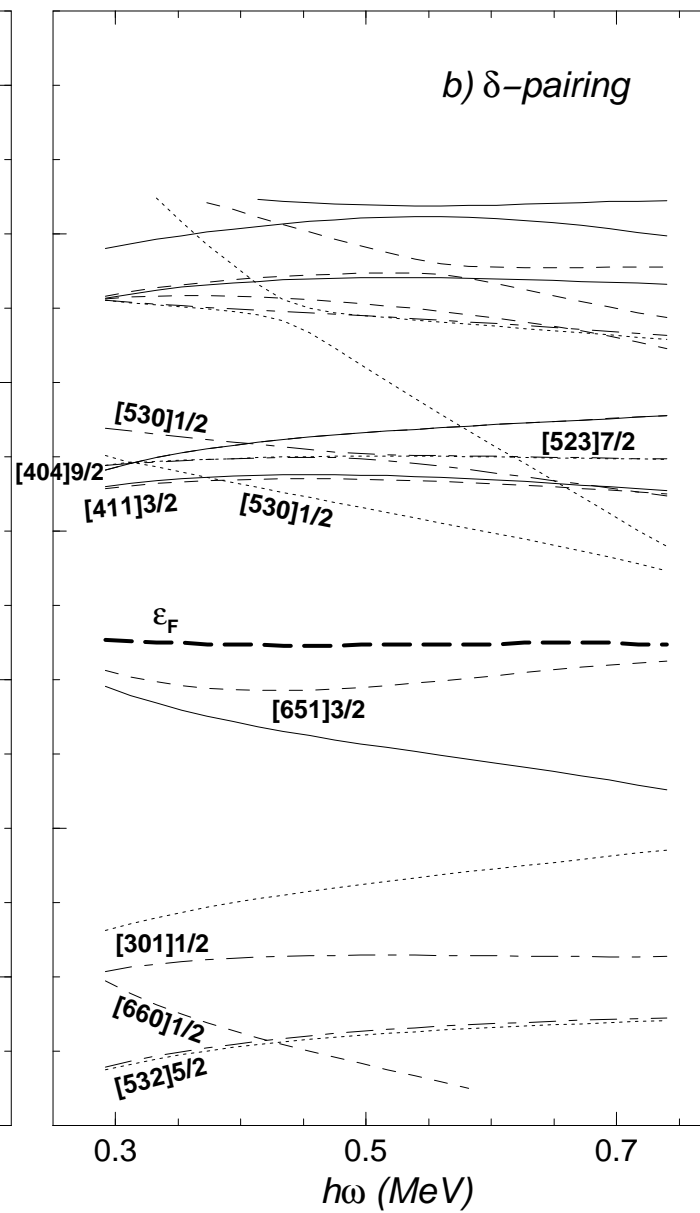
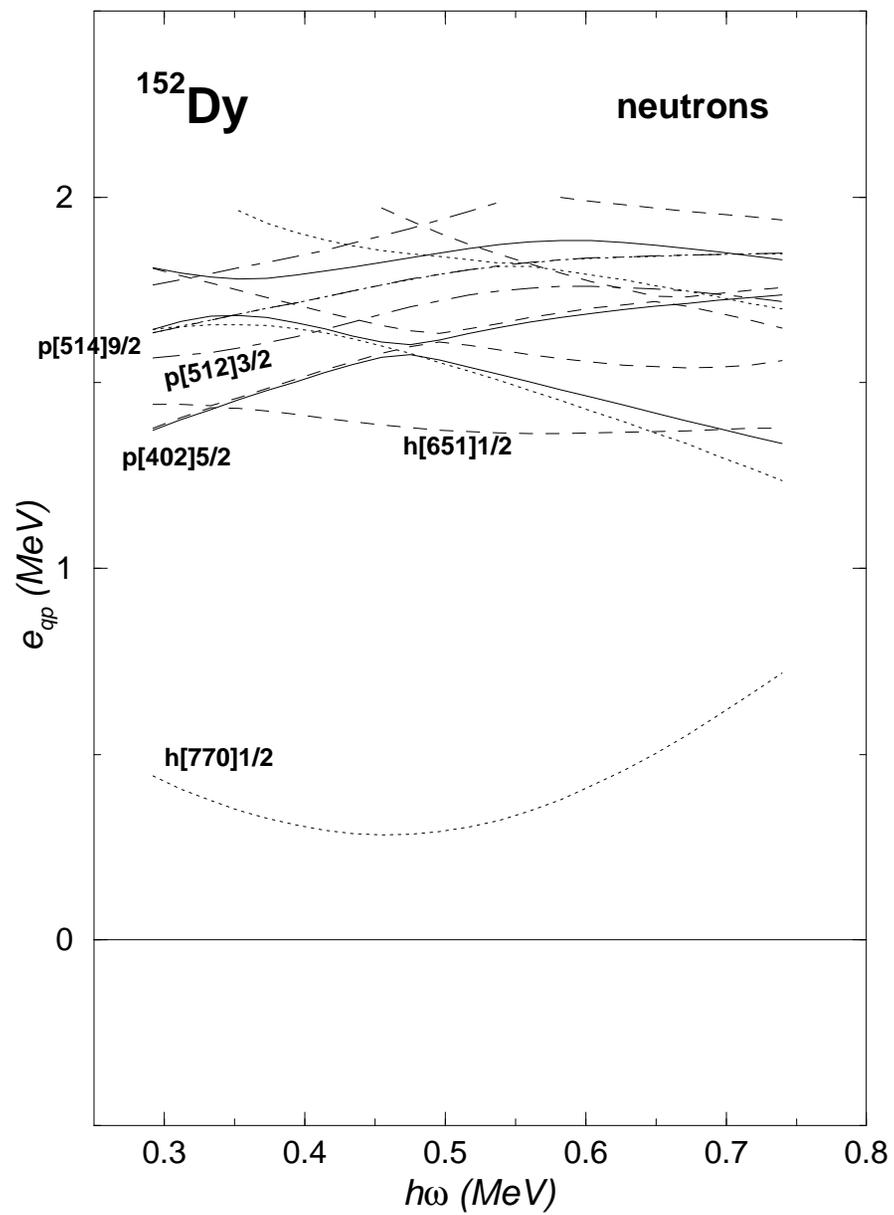
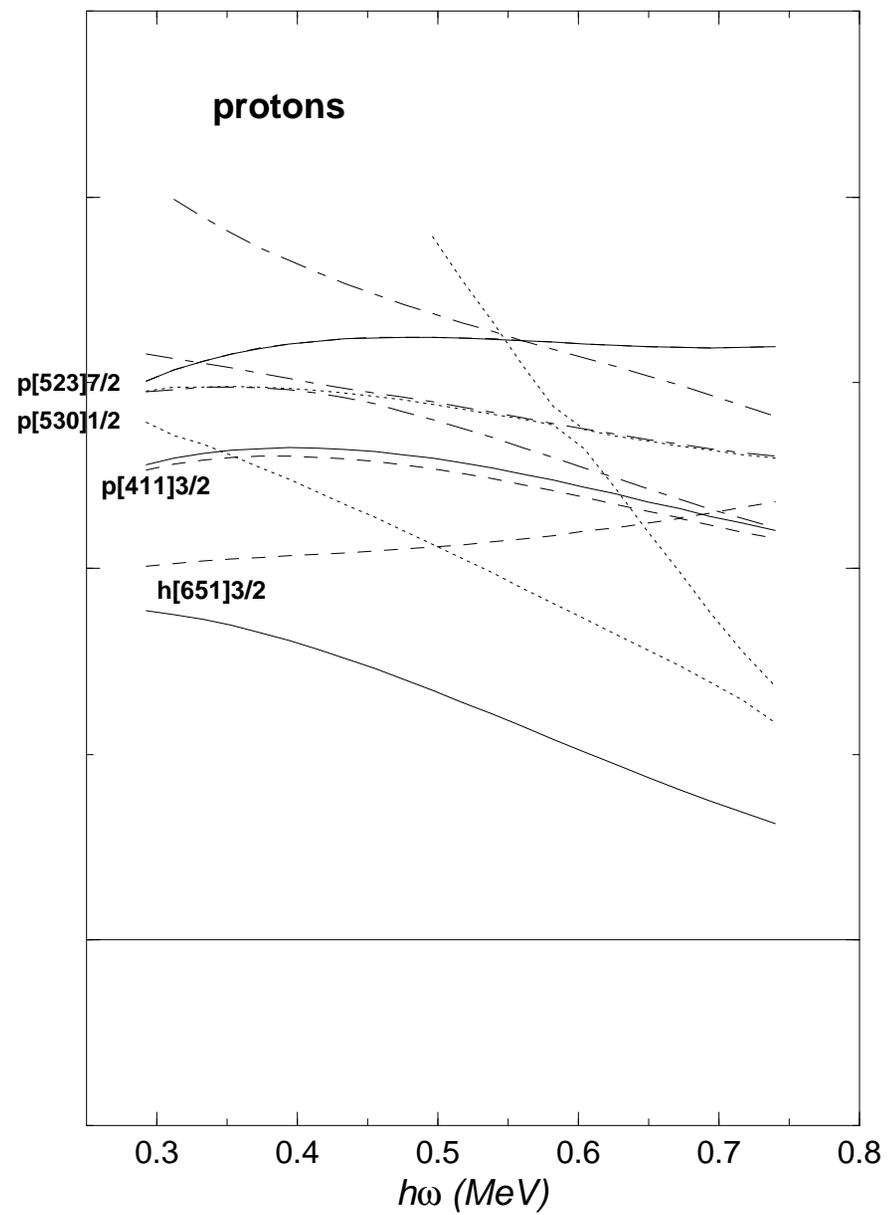
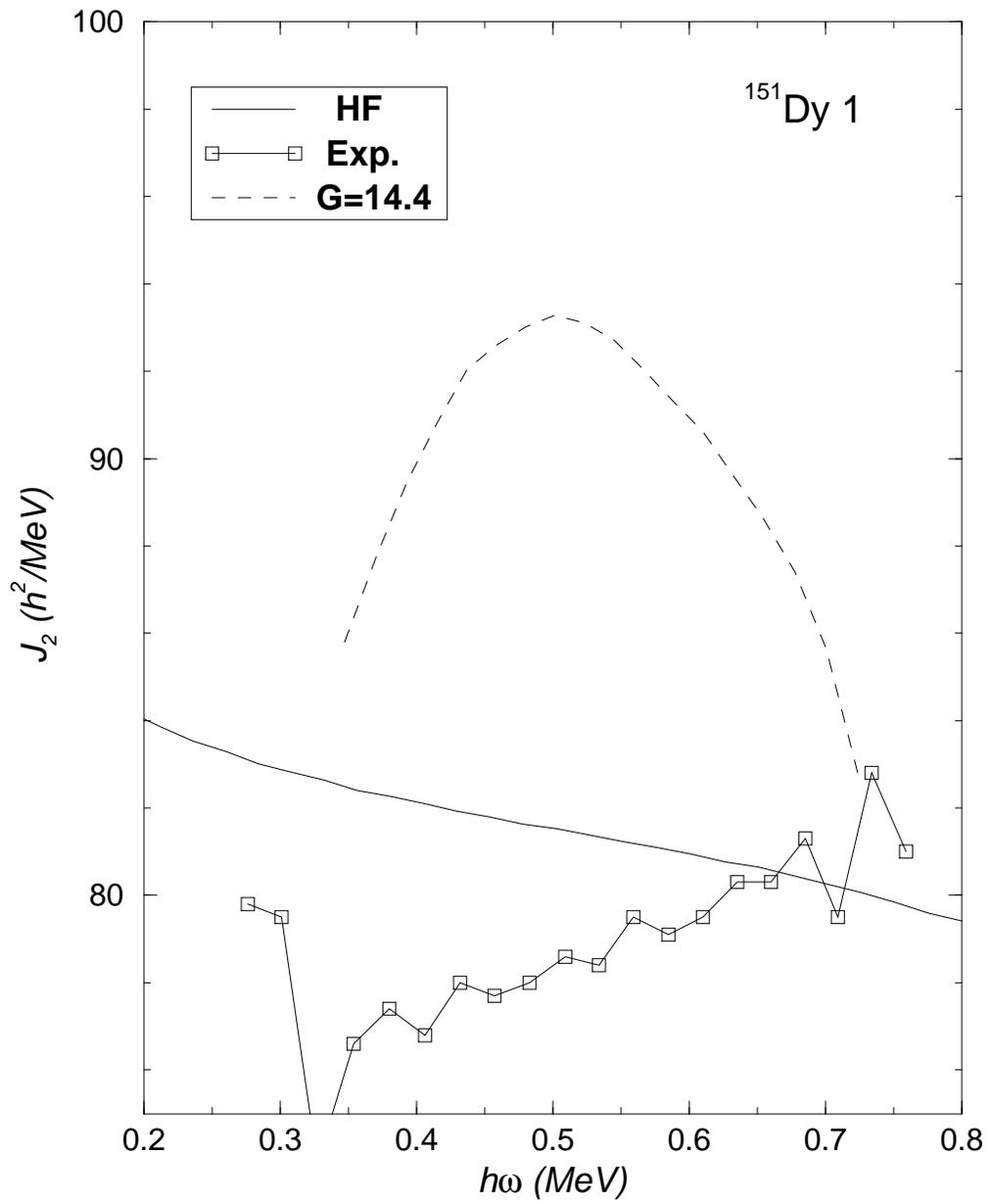
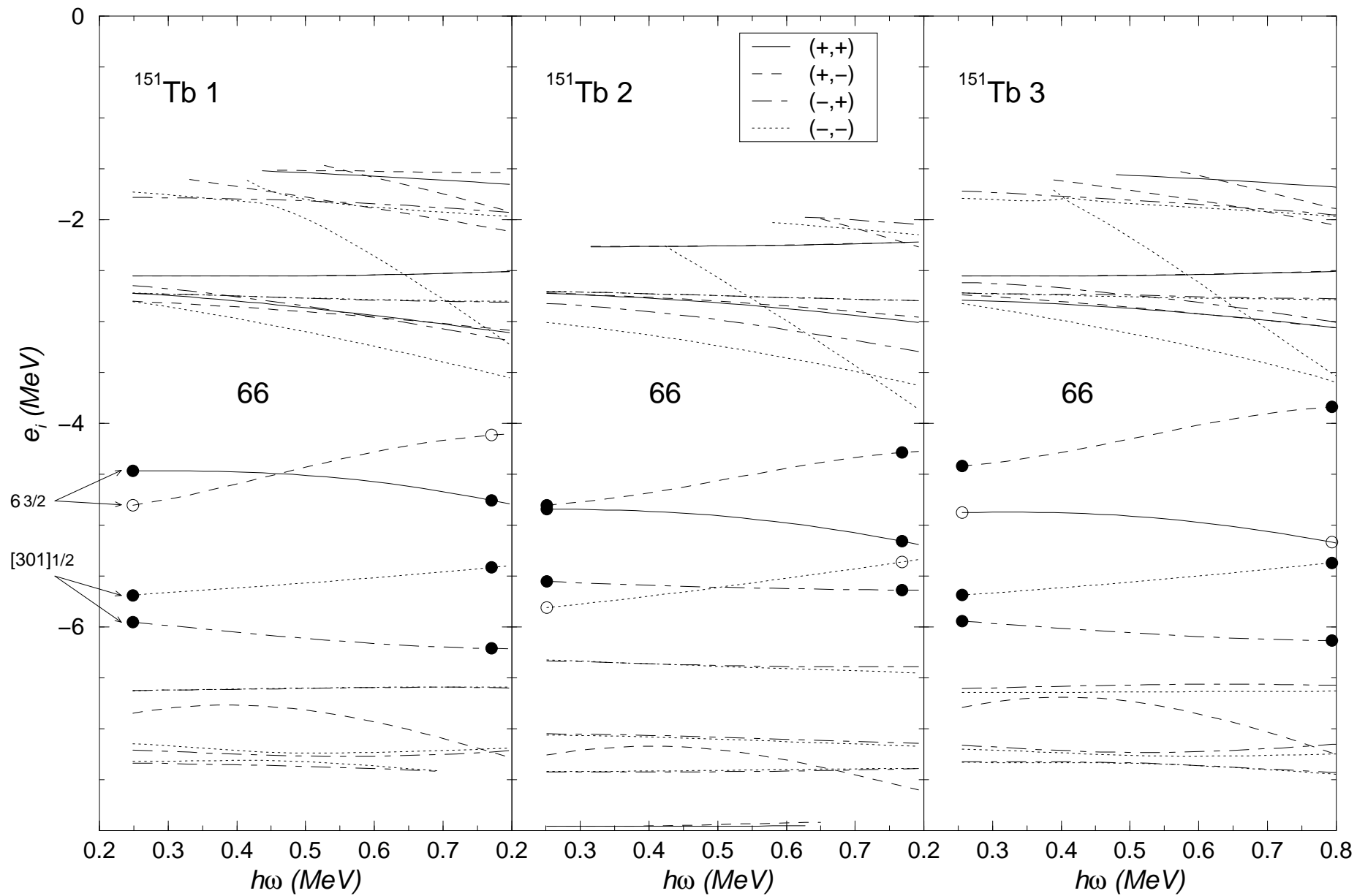
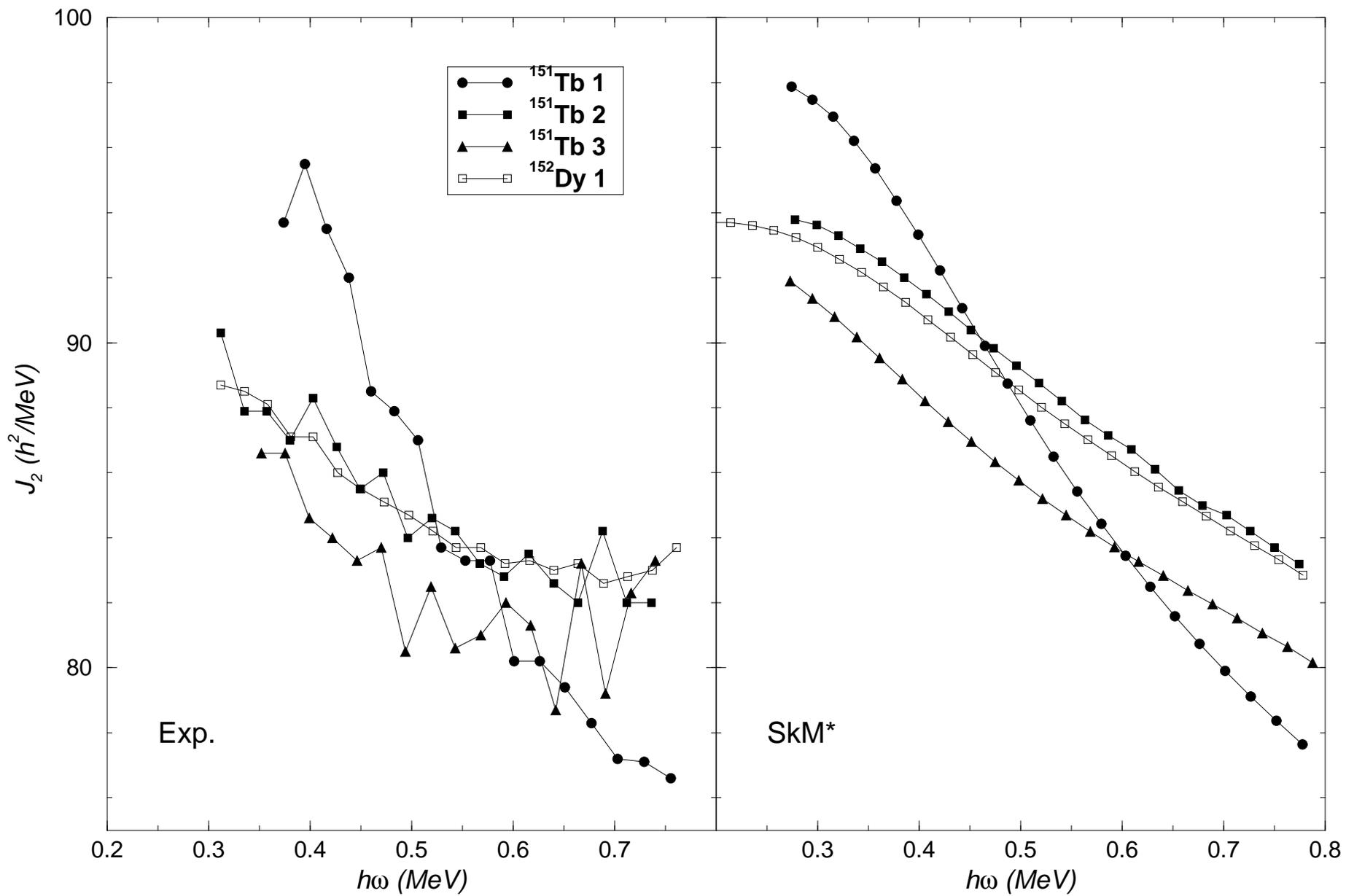
\end{document}